\documentclass[12pt,draft]{article}
\begin{document}
\title{Reply to "Comment on "Order parameter of A-like phase of $^3$He in aerogel""(cond-mat/0502549).}
\author{I.A.Fomin\\
P. L. Kapitza Institute for Physical Problems, \\
ul. Kosygina 2, 119334 Moscow,Russia}
\date{ }
\maketitle
\begin{abstract}
 The argument of V.P.Mineev and M.E.Zhitomirsky is based on
 unjustified omission of contribution of fluctuations to the
 free energy of superfluid $^3$He in aerogel.
 \end{abstract}
 \bigskip
% \pacs {PACS numbers: 67.57.Pq, 67.57.Bc, 67.57.De }

%\section {}
The random tensor field  $\eta_{jl}({\bf r})$ introduced in
references [5-7] of the "Comment" gives rise to spacial
fluctuations of the order parameter (in what follows references of
the "Comment" are prefixed by a capital C, e.g. [C5-C7]).
Coordinate dependent order parameter can be represented as a sum:
$A_{\mu j}({\bf r})=\bar{A}_{\mu j}+a_{\mu j}({\bf r})$, where
$\bar{A}_{\mu j}\neq 0$ is the average value of the order
parameter. Free energy of superfluid $^3$He can be also split in
two parts:
$$
F=F_0(\bar{A}_{\mu j})+F_{fl}(\bar{A}_{\mu j},a_{\mu j}),
$$
where $F_0(\bar{A}_{\mu j})$ is obtained by a substitution of
$\bar{A}_{\mu j}$ in the Ginzburg and Landau functional. The
remaining part of the free energy $F_{fl}$ contains averages of
the fluctuating increments $\overline{a_{\mu j}a_{\nu l}}$ etc..
Mineev and Zhitomirsky ignore the existence of $F_{fl}$ and
following the argument of Imry and Ma [C4] and Volovik [C3] arrive
at the conclusion that the quasi-isotropic A-like phases proposed
in Ref. [C5-C7] are energetically unfavorable. This conclusion is
erroneous since contribution of fluctuations to the free energy
can not be neglected as one can see from comparison of possible
gains of energy coming from the two parts.

Gain in the first part of energy $F_0(\bar{A}_{\mu j})$ is due to
the adjustment of orientation of the order parameter to
fluctuations of the random field on an optimal scale $L$, which
was estimated for $^3$He-A in Ref. [C3]. In terms of the tensor
$\eta_{jl}$ it is $L\sim \xi_0/\eta^2$ where $\xi_0$ is the
correlation length in the Cooper paired state. The corresponding
energy gain is $\delta F_0 \sim N(0)\tau T_c^2\eta^4$, where N(0)
is the density of states and $\tau=(T_c-T)/T_c$. To simplify the
argument the fact is used that for 96-98\% aerogel correlation
length of the random field $\xi_a$ is of the order of $\xi_0$ and
the ratio of two length is dropped.

When evaluating contribution of fluctuations one has to make
distinction between fluctuations in directions of degeneracy of
the order parameter (i.e. increments in these directions do not
change the free energy) and in the directions orthogonal to these.
Contribution of the latter fluctuations does not differ
essentially from that for conventional superconductors with a
scalar order parameter $\psi$. Representing $\psi$ as a sum of the
average order parameter $\bar\psi$ and fluctuating increment
$\zeta$ and using  result of Larkin and Ovchinnikov \cite{Larkin}
$\bar{\zeta^2}\sim\bar\psi^2(\eta^2/\sqrt\tau)$ one can estimate
the corresponding contribution to the free energy as $\delta
F_{\bot}\sim N(0)\tau^{3/2}T_c^2\eta^2$. This contribution is
already greater then $\delta F_0$ to the extent $\sqrt\tau/\eta^2$
($\eta^2/\sqrt\tau\ll 1$ is requied for fluctuations to remain
small in comparison with the average value of the order
parameter).

 Contribution to the free energy of fluctuations in directions of
 degeneracy of the order parameter contains an extra large factor
 $\sim \frac{1}{\xi_0}\int\frac{dk}{k^2}$ where $k$ is a wave
 vector of  corresponding fluctuations. The diverging on a low
 limit integral has to be cut on a scale $l_{\Delta}$ determined by
 the condition that the fluctuations of the order parameter are
 comparable with the order parameter itself i.e.
 $l_{\Delta}\sim\xi_0(\sqrt\tau/\eta^2)$. That brings an extra
 large factor $\sqrt\tau/\eta^2$ in the energy of fluctuations.
 As a result this energy competes with the fourth order terms in
 $F_0(\bar{A}_{\mu j})$ and an equilibrium order parameter is not
 necessarily a minimum of $F_0(\bar{A}_{\mu j})$. Criterion,
 formulated in Refs. [C5-C7] is a condition of turning to zero of
 coefficients in front of the diverging integrals. That gives much
 greater (by a factor of $\tau/\eta^4$) gain then that in
 $F_0(\bar{A}_{\mu j})$ discussed in the "Comment".
 Application of the criterion Refs. [C5-C7] selects
 a family of quasi-isotropic A-like phases which are proposed as 
 zero-order approximation for the observed A-like phase.

 For the particular example of a vector order parameter considered
 by Imry and Ma [C4] the diverging terms can be excluded only by
 turning to zero of the average value of the order parameter in
 agreement with their original statement.

 At the strength of the above estimations the argument presented in
 the "Comment" of Mineev and Zhitomirsky gives no reason for a
 revision of results of Refs.[C5-C7].

 I thank E.I.Kats and D.E.Khmelnitskii for useful discussion and
 the Institute of Laue and Langevin  for  hospitality and support
 of my visit to Grenoble.

  \end{document}